\RequirePackage{fix-cm}
\documentclass[smallextended]{svjour3}       
\smartqed  
\usepackage{graphicx}
\begin{document}

\title{2D dilute Bose mixture at low temperatures
}


\author{Pavlo Konietin     \and
        Volodymyr Pastukhov 
}


\institute{P.~Konietin \and V.~Pastukhov \at
              Department for Theoretical Physics, Ivan Franko National University of Lviv, 12 Drahomanov Street, Lviv-5, 79005, Ukraine\\
              \email{volodyapastukhov@gmail.com}           
           }

\date{Received: date / Accepted: date}

\maketitle

\begin{abstract}
The thermodynamic and superfluid properties of the dilute two-dimensional binary Bose mixture at low temperatures are discussed. We also considered the problem of the emergence of the long-range order in these systems. All calculations are performed by means of celebrated Popov's path-integral approach for the Bose gas with a short-range interparticle potential.
\end{abstract}

\section{Introduction}
\label{intro}
The spatial dimensionality plays a crucial role in the behavior of interacting many-boson systems. Perhaps, the most exciting phase diagram is obtained in the two-dimensional case (for review, see \cite{Posazhennikova,Hadzibabic}), where the ground-state Bose condensate state \cite{Schick,Lozovik,Ovchinnikov,Cherny,Pilati,Mazzanti} is altered by the low-temperature Beresinskii-Kosterlitz-Thouless (BKT) phase with the characteristic power-law \cite{Popov_72} decay of the one-particle density matrix. To describe these systems  appropriately one needs to extend \cite{Mora} the standard approach with the separated condensate and to use the phase-density formulation \cite{Popov}, renormalization-group \cite{Fisher,Kolomeisky,Dupuis_1,Dupuis_2,Rancon,Krieg} or the effective field-theoretic \cite{Andersen,Chien} treatments. Particularly Popov's theory allows to find out the low-energy structure of one-particle Green's functions \cite{Popov_Seredniakov} and improved version of this approach \cite{Andersen_etal,Khawaja} which takes into account phase fluctuations exactly is capable to explain \cite{Cockburn} experiments with two-dimensional Bose gases. 
In contrast to the two-dimensional Bose systems in an optical lattice \cite{Kagan,Altman,Kuklov1,Kuklov2,Schmidt,Guglielmino,Rousseau} that are characterized by nontrivial phase diagrams the equilibrium properties of the homogeneous binary 2D bosonic gases with continuous translational symmetry are less studied. Reliable results for the stability condition of the low-density mixtures were obtained in Refs.~\cite{Kolezhuk,Lee} by means of a renormalization-group technique and recently in  \cite{Petrov} the ground-state behavior of the two-dimensional binary Bose gases in the Bogoliubov approximation was discussed in a context of the droplet-formation phenomenon.

In the present paper we have analysed the ground-state properties of the two sorts of Bose particles interacting with the analytically tractable short-range two-body potentials in two spatial dimensions.
\section{Formulation}
\label{sec: 2}
We adopt a path-integral formulation for the two-component Bose system with the Euclidean action
\begin{eqnarray}\label{S}
	S=S_0+S_{int},
\end{eqnarray}
where the ideal-gas term reads
\begin{eqnarray}\label{S_0}
	S_0=\int dx \,\Psi^*_a(x)\left\{\partial_{\tau}+\frac{\hbar^2 }{2m_a}\Delta+\mu_a\right\}\Psi_a(x)
\end{eqnarray}
and the second one describes intra and interspecies two-body interaction
\begin{eqnarray}
	S_{int}=-\frac{1}{2}\int dx\int dx'\Phi_{ab}(x-x')
	|\Psi_a(x)|^2|\Psi_b(x')|^2.
\end{eqnarray}
The notations are typical: the $(2+1)$-vector $x=(\tau, {\bf r})$, $\int dx=\int^{\beta}_0d\tau\int_{\mathcal{A}}d {\bf r}$, where $\mathcal{A}$ is a large two-dimensional periodicity ``volume'', $\beta=1/T$ is the inverse temperature, the complex-valued $\beta$-periodic fields $\Psi_a(x)$ describe bosonic states, and the summation over repeated indices $a,b=(A,B)$ is assumed. We also denoted the chemical potentials $\mu_a$, masses of particles of each sort $m_a$ and interaction potentials $\Phi_{ab}(x)=\delta(\tau)\Phi_{ab}({\bf r})$. It is well-known that the Popov prescription \cite{Popov} is very convenient for the description of two-dimensional one-component many-boson systems, therefore in the remainder of the paper we applied this approach for the two-component Bose gas. The key idea is to introduce momentum scale $\hbar \Lambda$ that separates fields $\Psi_a(x)$ on the ``slowly'' $\psi_a(x)$ and ``rapidly'' $\tilde{\psi}_a(x)$ varying parts:
\begin{eqnarray}\label{Psi}
	\Psi_a(x)=\psi_a(x)+\tilde{\psi}_a(x), \,
	\Psi^*_a(x)=\psi^*_a(x)+\tilde{\psi}^*_a(x),
\end{eqnarray}
(note that $\int_{\mathcal{A}}d{\bf r}\,\psi_a(x)\tilde{\psi}_a(x)=0$) with the following functional integration over $\tilde{\psi}_a(x)$. In such a way after passing to the phase-density representation of the ``slowly'' varying $\psi_a(x)$ fields
\begin{eqnarray}\label{psi}
	\psi_a(x)=\sqrt{n_a(x)}e^{i\varphi_a(x)}, \,
	\psi_a^*(x)=\sqrt{n_a(x)}e^{-i\varphi_a(x)},
\end{eqnarray}
one obtains the effective hydrodynamic action which accurately captures the low-energy physics of the two-dimensional Bose systems in the whole temperature region. In the following we will discuss the zero-temperature limit only. If in addition the system is dilute, the role of the $\tilde{\psi}_a(x)$-fields is reduced to the replacement of the original interparticle potentials $\Phi_{ab}({\bf r})$ in the hydrodynamic action by the elements of the $t$-matrix (see Appendix for details). The final hydrodynamic action reads
\begin{eqnarray}\label{S_h}
	S_h=\int
	dx\left\{n_a(x)i\partial_{\tau}\varphi_a(x)-\frac{\hbar^2}{2m_a}n_a(x)(\nabla
	\varphi_a(x))^2\right.\nonumber\\
	\left.-\frac{\hbar^2}{8m_a}\frac{(\nabla n_a(x))^2}{n_a(x)}
	-\frac{1}{2}t_{ab}n_a(x)n_b(x)+\mu_a n_a(x)\right\},
\end{eqnarray}
where $t_{ab}$ is responsible for the two-body collisions. For the spatial homogeneous systems one makes use of the following decomposition in terms of the Fourier harmonics
\begin{eqnarray}\label{n_a}
	n_a(x)=n_a+\frac{1}{\sqrt{\mathcal{A}\beta}}\sum_{K}e^{iKx}n^a_{K}, \ \
	\varphi_a(x)=\frac{1}{\sqrt{\mathcal{A}\beta}}\sum_{K}e^{iKx}\varphi^a_{K},
\end{eqnarray}
where $K=(\omega_k, {\bf
	k})$ stands for the bosonic Matsubara frequency $\omega_k$ and two-dimensional wave-vector ${\bf k}$ (recall that $|{\bf k}|\le\Lambda$ and ${\bf k}\neq 0$). From the identities $-\partial \Omega/\partial\mu_a=n_aV$ and non-participation of the ``rapidly'' varying fields in the zero-temperature thermodynamics it is easy to show \cite{Pastukhov_q2D} that $n_a$ are equilibrium densities of two components. In the extremely dilute limit the properties of the system are correctly described by the Gaussian part of the action (\ref{S_h})
\begin{eqnarray}\label{S_G}
	S_{G}=-\beta\mathcal{A}\frac{t_{ab}}{2}n_an_b-\frac{1}{2}\sum_{K}
	\left\{\vphantom{\left[\frac{\varepsilon_a(k)}{2 n_a}\delta_{ab}
		+t_{ab}\right]}\omega_k\varphi^a_K n^a_{-K}-\omega_k\varphi^a_{-K}n^a_{K}
	\right.\nonumber\\
	\left.+2\varepsilon_a(k)n_a
	|\varphi^a_{K}|^2+\left[\frac{\varepsilon_a(k)}{2 n_a}\delta_{ab}
	+t_{ab}\right]n^a_{K}n^b_{-K}\right\},
\end{eqnarray}
(where $\varepsilon_a(k)=\hbar^2k^2/2m_a$ are the free-particle dispersions; $\delta_{ab}$ is the Kronecker delta) and performing a simple integration of the partition function in the low-temperature limit we obtained the ground-state energy of the binary Bose gas
\begin{eqnarray}\label{E_0}
	\frac{E_0}{\mathcal{A}} = \frac{1}{2}n_an_bt_{ab} +\frac{1}{2\mathcal{A}}\sum_{|{\bf k}|\le\Lambda }\left[E_{+}(k) + E_{-}(k)-\varepsilon_A(k)-\varepsilon_B(k)-n_at_{aa}\right].
\end{eqnarray}
Here the two branches \cite{Vakarchuk1,Vakarchuk2,Rovenchak} of the Bogoliubov spectrum read
\begin{eqnarray*}
	&&E^2_{\pm}(k) = \frac{1}{2}\left\{E^2_A(k) + E^2_B(k)\right.\nonumber\\
	&&\left. \pm \sqrt{[E^2_A(k) - E^2_B(k)]^2 +16\varepsilon_A(k)\varepsilon_B(k)n_An_Bt^2_{AB}}\right\},
\end{eqnarray*}
where $E^2_a(k)=\varepsilon^2_a(k)+2\varepsilon_a(k)n_at_{aa}$ represents the dispersion relation of an individual component. In the long-length limit these two branches of the spectrum of collective modes exhibit phonon-like behavior $E_{\pm}(k\rightarrow 0)=\hbar kc_{\pm}$. The terms presented in the last row of Eq.~(\ref{E_0}) do not appear during functional integration and should be inserted by hand \cite{Chang}. This procedure, however, is in agreement with calculations performed in the operator formalism \cite{Pastukhov_InfraredStr} and various regularization schemes \cite{Salasnich_Toigo}.

In order to study the superfluid properties of the two-component system let us suppose the each constituent to move slowly with velocity ${\bf v}_a$. In practice, while describing the superfluid hydrodynamics the smallness of ${\bf v}_a$ means that $|{\bf v}_a|\ll c_{-}$ providing the local thermodynamic equilibrium. The action of the moving Bose mixture is readily written by using the gauge transformation $\Psi_a(x)\rightarrow \Psi_a(x)e^{-im_a{\bf r}{\bf v}_a/\hbar}$ of the initial one (\ref{S})
\begin{eqnarray}\label{S_v}
	S_v=S-\beta\mathcal{A}\frac{\rho_av^2_a}{2}+i\hbar\int dx \Psi^*_a(x){\bf v}_a\nabla\Psi_a(x),
\end{eqnarray}
where $\rho_a$ are introduced for the mass densities. In the same manner as was argued above it is easy to show that the fields $\tilde{\psi}_a(x)$ do not contribute to the macroscopic properties of the two-component Bose system at low temperatures. Therefore the last term in (\ref{S_v}) only shifts the Matsubara frequencies $\omega_k\rightarrow \omega_k+i\hbar{\bf k}{\bf v}_a$ standing next to $\varphi^a_K n^a_{-K}$ term in the effective hydrodynamic action (\ref{S_h}), (\ref{S_G}). Again the explicit calculations for the energy of the moving Bose mixture can be made to the end only in the extremely dilute limit, where one can neglect the anharmonic terms in $S_h$. Up to the quadratic order over velocities ${\bf v}_a$ we have
\begin{eqnarray}\label{E_v}
	E_v/\mathcal{A}=E_0/\mathcal{A}+\frac{1}{2}\rho_{ab}{\bf v}_a{\bf v}_b+\ldots,
\end{eqnarray}
where the symmetric matrix of superfluid densities reads
\begin{eqnarray}\label{rho_ab}
	\rho_{ab}=\rho_a\delta_{ab}-\Delta\rho_{ab},
\end{eqnarray}
with 
\begin{eqnarray}\label{rho_AA}
	\Delta\rho_{AA}=\frac{1}{2\mathcal{A}\beta}\sum_K\frac{\hbar^2k^2[\omega^2_k+E^2_B(k)]}
	{[\omega^2_k+E^2_{+}(k)][\omega^2_k+E^2_{-}(k)]}\nonumber\\
	\times\left\{1-\frac{2\omega^2_k[\omega^2_k+E^2_B(k)]}
	{[\omega^2_k+E^2_{+}(k)][\omega^2_k+E^2_{-}(k)]}\right\},
\end{eqnarray}
\begin{eqnarray}\label{rho_AB}
	\Delta\rho_{AB}=-\frac{4}{\mathcal{A}\beta}\sum_K\frac{\hbar^2k^2\omega^2_k\varepsilon_A(k) \varepsilon_A(k)n_An_Bt^2_{AB}}{[\omega^2_k+E^2_{+}(k)]^2[\omega^2_k+E^2_{-}(k)]^2},
\end{eqnarray}
calculated in two dimensions. At zero temperature the whole system is superfluid and the Galilean invariance requires $\Delta\rho_{AA}=\Delta\rho_{BB}=-\Delta\rho_{AB}=\Delta\rho$ that can be easily verified by the direct integration over the Matsubara frequencies in Eqs.~(\ref{rho_AA}), (\ref{rho_AB}).

In two-dimensional systems of bosons residing exactly in the ground state the lowest one-particle state is macroscopically occupied. Nevertheless the developed thermodynamic fluctuations at any finite temperatures totally deplete this Bose condensate, it is interesting to calculate its value for the weakly-interacting gas from the methodological point of view. In the Popov approach the condensate density is obtained as follows:
\begin{eqnarray}
	\sqrt{n_{0a}}=\langle\Psi_a(x)\rangle|_{T=0}\rightarrow \langle\psi_a(x)\rangle,
\end{eqnarray}
where the last average should be calculated very carefully \cite{Pastukhov_twocomp} within the hydrodynamic action (\ref{S_h})
\begin{eqnarray}
	\sqrt{n_{0a}}=\lim_{\tau'\rightarrow \tau-0}\langle
	\sqrt{n_a(x)}e^{i\varphi_a(x')}\rangle{\big|}_{{\bf r}'={\bf r}}.
\end{eqnarray}
Taking into account the Gaussian fluctuations only, one gets the result
\begin{eqnarray}
	n_{0A}=n_{A}-\frac{1}{2
		\mathcal{A}}\sum_{|{\bf k}| \le \Lambda}\left\{
	\frac{\varepsilon_A(k)+n_At_{AA}}{E_{+}(k)+E_{-}(k)}\left[1+\frac{E^2_B(k)}{E_{+}(k)E_{-}(k)}\right]-1 \right\}.
\end{eqnarray}

In addition to the superfluid properties of the binary Bose mixture the quantity $\rho_{ab}$ together with the inverse compressibility matrix $\partial \mu_a/\partial n_b$ determine two velocities of a sound propagation \cite{Andreev} in the two-component bosonic medium. Finally it also identifies the exponents of the one-body density matrices
\begin{eqnarray}
	F_{ab}(|{\bf r}-{\bf r}'|)=\langle\Psi^*_a(x)\Psi_b(x')\rangle|_{\tau\rightarrow \tau'},
\end{eqnarray}
at large particle separations when $T\neq0$. Particularly by using results of Ref.~\cite{Pastukhov_twocomp} for the various two-legged vertices it is easy to argue that the {\it exact} asymptotic behavior of $F_{ab}(r)$ reads
\begin{eqnarray}
	F_{ab}(r\rightarrow \infty)\propto \frac{\delta_{ab}}{r^{\eta_a}}, \ \ \eta_a=\frac{m^2_aT}{2\pi \hbar^2}\rho^{-1}_{aa},
\end{eqnarray}
that indicates the Berezinskii-Kosterlitz-Thouless phase (here $\rho^{-1}_{ab}$ are elements of the inverse to $\rho_{ab}$ matrix). At very low temperatures when the temperature depletion of the superfluid density which is of order $T^3$ for the two-dimensional systems can be neglected the exponents $\eta_a$ are fully determined by $\Delta\rho$ given by Eq.~(\ref{rho_AA}) for the dilute mixtures.

\section{Model with the short-range interaction}
\label{sec:3}
For the specific calculations we choose the two-body potentials in the following form
\begin{eqnarray}\label{Phi_r}
	\Phi_{ab}({\bf r})=\frac{g_{ab}}{\pi R^2_{ab}}e^{-r^2/R^2_{ab}},
\end{eqnarray}
where $g_{ab}$ are the coupling constants responsible for the interaction strength and parameters $R_{ab}$ characterize the effective range of the potentials. In the limit $R_{ab}\rightarrow 0$ the function in Eq.~(\ref{Phi_r}) tends to $\delta$-function and the coupling constants $g_{ab}$ to the leading order can be rewritten via the experimentally measured $s$-wave scattering lengths $l_{ab}$ \cite{Volosniev}
\begin{eqnarray}\label{g_ab}
	\frac{1}{g_{ab}}=-\frac{m_{ab}}{\pi\hbar^2}\ln[e^{\gamma/2}l_{ab}/R_{ab}],
\end{eqnarray}
where we denote the reduced masses $1/m_{ab}=1/m_a+1/m_b$ and $\gamma=0.57721\ldots$ is the Euler-Mascheroni constant. On the other hand, by using equations obtained in Appendix we are in position to express $g_{ab}$ via elements of the $t$-matrix
\begin{eqnarray}
	\frac{1}{g_{ab}}=\frac{1}{t_{ab}}-\frac{m_{ab}}{\pi\hbar^2}\ln[2e^{-\gamma/2}/(R_{ab}\Lambda)],
\end{eqnarray}
found in the limit $R_{ab}\Lambda\ll 1$. These two equation allow to eliminate the dependence on the nonuniversal parameters $g_{ab}$ and $R_{ab}$ in the formula for the matrix $t_{ab}$ and as a consequence in the hydrodynamic action (\ref{S_h})
\begin{eqnarray}\label{t_ab}
	t_{ab}=\frac{\pi \hbar^2}{m_{ab}}\frac{1}{\ln\left[\frac{2e^{-\gamma}}{l_{ab}\Lambda}\right]}.
\end{eqnarray}
To simplify further consideration we assume the equal-mass limit $m_a=m_b=m$ in which the excitation spectrum of the system looks Bogoliubov-like $E_{\pm}(k)=\hbar^4k^4/4m^2+\hbar^2k^2c^2_{\pm}$ with the sound velocities given by
\begin{eqnarray}\label{c_pm}
	c^2_{\pm}=(c^2_A+c^2_B)/2\pm \sqrt{(c^2_A-c^2_B)^2/4+n_An_Bt^2_{AB}/m^2},
\end{eqnarray}
where $c_a=\sqrt{n_at_{aa}/m}$ denote the sound velocities of individual components. A great advantage of two spatial dimensions in the equal-mass limit is that all integrals can be performed analytically to the very end. Particularly, for the ground-state energy (\ref{E_0}) one obtains
\begin{eqnarray}\label{E_Bog}
	E_{0}/\mathcal{A}=\frac{1}{2}t_{ab}n_an_b+\frac{m^3}{4\pi\hbar^2}\sum_{j=\pm}c^4_j\ln\left[e^{1/4}\frac{mc_j}{\hbar\Lambda}\right],
\end{eqnarray}
(recall that this expression is valid only in the limit $\hbar^2\Lambda^2/m\gg \mu_{a}\sim t_{ab}n_b$) which is consistent with the general equation for two-dimensional binary systems \cite{Werner}. With the same accuracy we have calculated the quantity $\Delta\rho$ determining the matrix of superfluid densities
\begin{eqnarray}
	\Delta\rho/m=\frac{n_An_Bt^2_{AB}}{8\pi\hbar^2}\frac{c^4_{+}+2c^2_{+}c^2_{-}
		\ln[c^2_{-}/c^2_{+}]-c^4_{-}}{(c^2_{+}-c^2_{-})^3},
\end{eqnarray}
and in our approximation shifts the BKT exponents $\eta_a=\frac{mT}{2\pi\hbar^2 n_a}\left[1+\Delta\rho/(mn_a)+\ldots\right]$ at low temperatures. For the dilute Bose mixture with the symmetric interaction $t_{AA}=t_{BB}$ at any density ratios this effect becomes more tangible when the system is extremely close to the phase-separation region. The interaction-induced
condensate depletion of a component $A$ at absolute zero reads
\begin{eqnarray}\label{n_Bog}
	n_A-n_{0A}=\frac{m^2}{4\pi\hbar^4}\left\{\vphantom{\ln\left[\frac{c^2_{-}}{c^2_{+}}\right]}c^2_A+\frac{(c^2_A-c^2_{+})(c^2_A-c^2_{-})}{c^2_{+}-c^2_{-}}\ln\left[\frac{c^2_{-}}{c^2_{+}}\right]\right\}.
\end{eqnarray}
The final stage of the above calculations is the determination of the cut-off parameter $\Lambda$. The most natural way to find it was proposed originally by Popov providing the minimization of the thermodynamic potential (in our zero-temperature case just the ground-state energy). It is easy to confirm by the direct calculations with a logarithmic accuracy that $E_{0}$ does not depend on $\Lambda$ \cite{Petrov}, i.e.,  $\partial E_{0}/\partial \Lambda=0$. This observation allows to choose the cut-off parameter  up to an irrelevant factor from the dimensional arguments, for instance, $\Lambda^2\sim \max\{n_A,n_B\}$ or $\Lambda^2\sim n=n_A+n_B$, which correctly reproduces the one-component limits \cite{Schick} and together with smallness of the coupling constants $g_{ab}$ provide the system not to be in crystalline phase \cite{Kroiss}.

It is instructive to apply the above-presented Popov's approach to the  one-component model. The properties of the single-component two-dimensional Bose gas can be obtained from Eqs.~(\ref{E_Bog}), (\ref{n_Bog}) by tending density of one sort of particles (let say $n_B$) to zero. Then identifying $n_A$ with the total density $n$ of the system and setting  $m_A=m$, $t_{AA}=t$ ($l_{AA}=l$) we obtain in the leading order that $\partial E_{0}/\partial \Lambda=0$, i.e., we can again choose parameter $\Lambda^2\sim n$ by using the dimensional arguments.  But the condition of cancellation of subleading terms gives for the ground-state energy $E_0/\mathcal{A}=\frac{2\pi\hbar^2n^2}{m}\lambda^2(1-\lambda^2/2)$, where $\Lambda^2=4\pi e n\lambda^2$ and $\lambda$ is determined by the transcendental equation $1/\lambda^2=\ln\frac{1}{nl^2}-1-2\gamma-\ln\pi+\ln\frac{1}{\lambda^2}$. The iterative solution in the dilute regime $nl^2\ll 1$ yields for the coefficient in formula for $E_0$: 
\begin{eqnarray}	
	\lambda^2(1-\lambda^2/2)\to \frac{1}{1/2+1/\lambda^2}= \frac{1}{\ln\frac{1}{nl^2}+\ln\ln\frac{1}{nl^2}-1/2-2\gamma-\ln\pi}+\ldots,
\end{eqnarray}	
which should be compared with that of Refs.~\cite{Mora09,Astrakharchik_etal}.

It is easily seen that the Popov's treatment correctly reproduces to the leading order the results of more sophisticated approaches and we therefore may use this formulation to obtain the beyond-mean-field stability condition of two-component systems. Firstly, of course, one should calculate the cut-off parameter, which is determined as follows
\begin{eqnarray}\label{cut_off}	
\sum_{j=\pm}\ln\left[e^{1/2}\frac{mc_j}{\hbar\Lambda}\right]\frac{\partial c^4_j}{\partial \Lambda}=0,
\end{eqnarray}	
in this case. The general consideration of the thermodynamic stability leads to complicated transcendental equation that has to be solved for an arbitrary set of $s$-wave scattering lengths and concentrations of ingredients. For very dilute systems, however, only region close to the mean-field phase-separation condition $\det t_{ab}=0$ (here $t_{ab}=4\pi\hbar^2/[m|\ln nl^2_{ab}|]$) is the most interesting. Thus assuming that all $l_{ab}$ are of the same order magnitude and again introducing dimensionless cut-off parameter $\lambda^2=\Lambda^2/(4\pi e n)$ we found the asymptotic solution of Eq.~(\ref{cut_off})  
\begin{eqnarray}	
	1/\lambda^2= \ln\frac{1}{nl^2}+\ln\ln\frac{1}{nl^2}-1-2\gamma-\ln\pi +\frac{n_an_b}{n^2}\ln\frac{l^2}{l^2_{ab}}+\ldots,
\end{eqnarray}	
where $l\sim l_{ab}$ is arbitrary length-scale and the appropriate ground-state energy 
\begin{eqnarray}	
E_0/\mathcal{A}=\frac{2\pi \hbar^2}{m}\frac{n_an_b}{\ln\frac{1}{nl^2_{ab}}+\ln\ln\frac{1}{nl^2_{ab}}-1/2-2\gamma-\ln\pi } +\ldots,
\end{eqnarray}	
calculated with the same accuracy. This result particularly states that the mean-field stability condition $l_{AA}l_{BB}>l^2_{AB}$ of the binary two-dimensional Bose mixtures is unaffected by the Bogoliubov approximation (recall that the obtained formulae are correct only for very dilute systems, i.e., when $1/|\ln nl^2_{ab}|\ll 1)$.

\section{Conclusions}
\label{sec:4}
In summary, by means of the Popov's prescription we have analyzed the equilibrium thermodynamic and superfluid properties of the two-dimensional dilute binary Bose mixtures. It is shown that the presence of the interspecies interaction shifts the exponents of one-particle density matrices in the Berezinskii-Kosterlitz-Thouless phase at low temperatures. Our findings could serve a starting platform for further theoretical study of the impact of the beyond-mean-field effects on the macroscopic properties of the two-component two-dimensional Bose systems. But in order to compare the obtained results with experiments one should extend the proposed approach to more realistic binary Bose mixtures, namely to the two-component systems loaded in the pancake trap. The second question has to be answered is the inclusion of nonuniversal parts of the interaction potentials which are recently found \cite{Salasnich} to affect significantly on the thermodynamic properties of two-dimensional bosons in the one-component case. All these problems will be considered in future studies.

\begin{acknowledgements}
We thank Dr.~A.~Rovenchak for invaluable suggestions.
This work was partly supported by Project FF-30F (No.~0116U001539) from the
Ministry of Education and Science of Ukraine.
\end{acknowledgements}

\section{Appendix}
\label{sec:5}
After transformation (\ref{Psi}) the action of the two-component Bose mixture is
	\begin{eqnarray}\label{S_trans}
		S\rightarrow\int dx \,\psi^*_a(x)\left\{\partial_{\tau}+\frac{\hbar^2 }{2m_a}\Delta+\mu_a\right\}\psi_a(x)\nonumber\\
		+\int dx \,\tilde{\psi}^*_a(x)\left\{\partial_{\tau}+\frac{\hbar^2 }{2m_a}\Delta+\mu_a\right\}\tilde{\psi}_a(x)\nonumber\\ 
		-\frac{1}{2}\int dx\int dx'\Phi_{ab}(x-x')|\psi_a(x)|^2|\psi_b(x')|^2\nonumber\\
		-\int dx\int dx'\Phi_{ab}(x-x')\left\{\psi^*_a(x)
		\psi^*_b(x')\psi_b(x')\tilde{\psi}_a(x)+{\rm c.c}\right\}\nonumber\\ 
		-\int dx\int dx'\Phi_{ab}(x-x')\left\{|\psi_a(x)|^2|\tilde{\psi}_b(x')|^2+\psi^*_a(x)
		\tilde{\psi}^*_b(x')\psi_b(x')\tilde{\psi}_a(x)\right\}\nonumber\\ 
		-\frac{1}{2}\int dx\int dx'\Phi_{ab}(x-x')\left\{\psi^*_a(x)\psi^*_b(x')\tilde{\psi}_b(x')\tilde{\psi}_a(x)+{\rm c.c}\right\}\nonumber\\-\int dx\int dx'\Phi_{ab}(x-x')\left\{\tilde{\psi}^*_a(x)
		\tilde{\psi}^*_b(x')\tilde{\psi}_b(x')\psi_a(x)+{\rm c.c}\right\}\nonumber\\
		-\frac{1}{2}\int dx\int dx'\Phi_{ab}(x-x')|\tilde{\psi}_a(x)|^2|\tilde{\psi}_b(x')|^2,
	\end{eqnarray}
where nine interaction vertices are presented diagrammatically in Fig.~1.
\begin{figure}[h!]
	\centerline{\includegraphics
		[width=0.5
		\textwidth,clip,angle=-0]{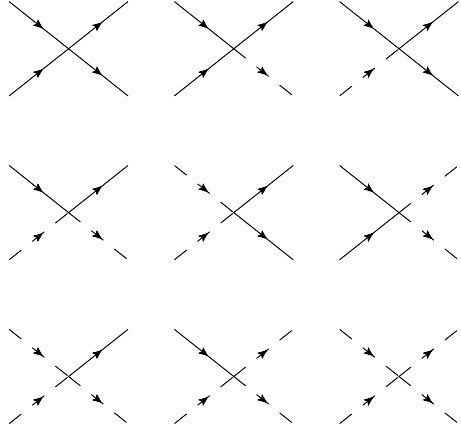}}
	\caption{Diagrammatic representation of the vertices appearing in Eq.~(\ref{S_trans}). The solid and dashed lines stand for the $\psi_a(x)$ and $\tilde{\psi}_a(x)$, respectively.}
	\label{fig:1} 
\end{figure}
For a very dilute system while integrating out $\tilde{\psi}$-fields one can use simple perturbation theory considering the ideal gas action (the second term in Eq.~(\ref{S_trans})) as a zero-order approximation. Furthermore, in the low-temperature limit the only nonzero contribution to the effective action governing ``slowly'' varying fields is given by the graphs depicted in Fig.~2.
\begin{figure}[h!]
	\centerline{\includegraphics
		[width=0.75
		\textwidth,clip,angle=-0]{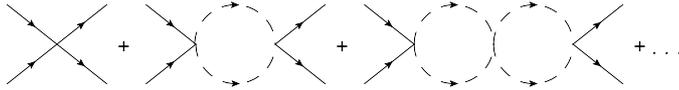}}
	\caption{Renormalization of the interaction potential in the hydrodynamic action.}
	\label{fig:2} 
\end{figure}
This infinite series of diagrams is summed up to give the linear integral equation for the renormalized symmetrical vertices
\begin{eqnarray}\label{t_eq}
		t_{ab}(P,Q|Q+K,P-K)=[g_{ab}(k)+\delta_{ab}g_{aa}(p-k-q)]/2^{\delta_{ab}}\nonumber\\
		-\frac{1}{2^{\delta_{ab}}\mathcal{A}\beta}\sum_{S}[g_{ab}(s)+\delta_{ab}g_{aa}(p-s-q)]\nonumber\\
		\times G_a(P-S)G_b(Q+S)t_{ab}(P-S,Q+S|Q+K,P-K),
\end{eqnarray}
where $g_{ab}(k)$ is the Fourier transform of $\Phi_{ab}({\bf r})$ and we used notation for Green's functions of the ideal gases $G_a(P)=1/[i\omega_p-\varepsilon_a(p)+\mu_a]$. In the dilute limit, where $\mu_{a}\ll \hbar^2\Lambda^2/m_a$ (the most natural choice in the two-component case is $\Lambda^2\sim n_A+n_B$) and the physically relevant region of the wave-vectors integration $p,q,k\sim \sqrt{m\mu_{a}}/\hbar\ll\Lambda$, we can neglect the dependence on $P$ and $Q$ under the integral in Eq.~(\ref{t_eq}). If it is also assumed that the interaction potentials are short-ranged, i.e., $g_{ab}(s)$ weakly depends on $s$ then the $t_{ab}(-S,S|0,0)$ is also independent on the transferred momentum. The latter observation allows to find the asymptotic solution of (\ref{t_eq}) (denoting $t_{ab}(0,0|0,0)\equiv t_{ab}$)
\begin{eqnarray}
	\frac{1}{t_{ab}}=\frac{1}{g_{ab}(0)}
	+\int^{\infty}_{\Lambda}\frac{ds\,s}{2\pi}\frac{g_{ab}(s)}{g_{ab}(0)}
	\frac{1}{\varepsilon_{a}(s)+\varepsilon_{b}(s)}.
\end{eqnarray}
Actually $t_{ab}$ enters the hydrodynamic action as a matrix of the effective coupling constants.


\begin{thebibliography}{99}
%
%
\bibitem{Posazhennikova} A.~Posazhennikova, Phys.~Rev.~Mod. {\bf 78},  1111 (2006).
\bibitem{Hadzibabic} Z.~Hadzibabic, and J.~Dalibard, Riv.~Nuovo~Cim. {\bf 34}, 389
(2011).

\bibitem{Schick} M.~Schick, Phys.~Rev.~A {\bf 3}, 1067 (1971).
\bibitem{Lozovik} Yu.~E.~Lozovik, V.~I.~Yudson, Physica~A {\bf 93}, 493 (1978).
\bibitem{Ovchinnikov} A.~A.~Ovchinnikov, J.~Phys.:~Condens.~Matter {\bf 5}, 8665 (1993).
\bibitem{Cherny} A.~Yu.~Cherny and A.~A.~Shanenko, Phys.~Rev.~E {\bf 64}, 027105 (2001).
\bibitem{Pilati} S.~Pilati, J.~Boronat, J.~Casulleras, and S.~Giorgini, Phys.~Rev.~A {\bf 71}, 023605 (2005).
\bibitem{Mazzanti} F.~Mazzanti, A.~Polls, and A.~Fabrocini, Phys.~Rev.~A {\bf 71}, 033615 (2005).


\bibitem{Popov_72} V.~N.~Popov, Theor.~Math.~Phys. {\bf 11}, 565 (1972).

\bibitem{Mora} C.~Mora and Y.~Castin, Phys.~Rev.~A {\bf 67}, 053615 (2003).

\bibitem{Popov} V.~N.~Popov, {\it Functional Integrals and Collective
	Excitations} (Cambridge University Press, Cambridge, 1987).


\bibitem{Fisher} D.~S.~Fisher and P.~C.~Hohenberg, Phys.~Rev.~B {\bf 37}, 4936 (1988).
\bibitem{Kolomeisky} E.~B.~Kolomeisky and J.~P.~Straley, Phys.~Rev.~B {\bf 46}, 11749 (1992).
\bibitem{Dupuis_1} N.~Dupuis, Phys.~Rev.~Lett. {\bf 102}, 190401 (2009).
\bibitem{Dupuis_2} N.~Dupuis, Phys.~Rev.~A {\bf 80}, 043627 (2009).
\bibitem{Rancon} A.~Rancon and N.~Dupuis, Phys.~Rev.~A {\bf 85}, 063607 (2012).
\bibitem{Krieg} J.~Krieg, D.~Strassel, S.~Streib, S.~Eggert, and P.~Kopietz, Phys.~Rev.~B {\bf 95}, 024414 (2017).

\bibitem{Andersen} J.~O.~Andersen, Eur.~Phys.~J.~B {\bf 28}, 389 (2002).
\bibitem{Chien} C.~C.~Chien, J.~H.~She, and F.~Cooper, Ann.~Phys. {\bf 347}, 192 (2014).

\bibitem{Popov_Seredniakov} V.~N.~Popov and A.~V.~Seredniakov, Sov.~Phys.~JETP {\bf 50},
193 (1979).

\bibitem{Andersen_etal} J.~O.~Andersen, U.~Al~Khawaja, and H.~T.~C.~Stoof, Phys.~Rev.~Lett. {\bf 88},  070407  (2002).
\bibitem{Khawaja} U.~Al~Khawaja, J.~O.~Andersen, N.~P.~Proukakis, and H.~T.~C.~Stoof, Phys.~Rev.~A {\bf 66}, 013615 (2002).

\bibitem{Cockburn} S.~P.~Cockburn and N.~P.~Proukakis, Phys.~Rev.~A {\bf 86}, 033610 (2012).

\bibitem{Kagan} M.~Yu.~Kagan and D.~V.~Efremov, Phys.~Rev.~B {\bf 65}, 195103 (2002).
\bibitem{Altman} E.~Altman, W.~Hofstetter, E.~Demler and M.~D.~Lukin, New~J.~Phys. {\bf 5}, 113 (2003).
\bibitem{Kuklov1} A.~Kuklov, N.~Prokof’ev, and B.~Svistunov, Phys.~Rev.~Lett. {\bf 92}, 050402 (2004)
\bibitem{Kuklov2} A.~Kuklov, N.~Prokof’ev, and B.~Svistunov, Phys.~Rev.~Lett. {\bf 92}, 030403 (2004).
\bibitem{Schmidt} K.~P.~Schmidt, J.~Dorier, A.~Lauchli, and F.~Mila, Phys.~Rev.~B {\bf 74}, 174508 (2006).
\bibitem{Guglielmino} M.~Guglielmino, V.~Penna, and B.~Capogrosso-Sansone, Phys.~Rev.~A {\bf 82}, 021601(R) (2010).
\bibitem{Rousseau} L.~de~Forges~de~Parny and V.~G.~Rousseau, Phys.~Rev.~A {\bf 95}, 013606 (2017).

\bibitem{Kolezhuk} A.~K.~Kolezhuk, Phys.~Rev.~A {\bf 81}, 013601 (2010).
\bibitem{Lee} Yu-Li~Lee, and Yu-Wen~Lee, J.~Phys.~Soc.~Jpn. {\bf 80}, 044003 (2011).

\bibitem{Petrov} D.~S.~Petrov and G.~E.~Astrakharchik, Phys.~Rev.~Lett. {\bf 117},  100401 (2016).



\bibitem{Pastukhov_q2D} V.~Pastukhov, Ann. Phys., {\bf 372}, 149 (2016).

\bibitem{Vakarchuk1} I.~O.~Vakarchuk, V.~S.~Pastukhov, J.~Phys.~Stud. {\bf 12}, 1001 (2008).
\bibitem{Vakarchuk2} I.~O.~Vakarchuk, V.~S.~Pastukhov, J.~Phys.~Stud. {\bf 12}, 3002 (2008).

\bibitem{Rovenchak} A.~Rovenchak, Low~Temp.~Phys. {\bf 42}, 36 (2016).

\bibitem{Chang} Chih-chun~Chang and R.~Friedberg, Phys.~Rev.~B {\bf 51}, 1117 (1995).

\bibitem{Pastukhov_InfraredStr} V.~Pastukhov, J.~Low Temp.~Phys. {\bf 186}, 148 (2017).

\bibitem{Salasnich_Toigo} L.~Salasnich and F.~Toigo, Phys.~Rep. {\bf 640}, 1 (2016).

\bibitem{Pastukhov_twocomp} V.~Pastukhov, Phys.~Rev.~A {\bf 95}, 023614 (2017).

\bibitem{Andreev} A.~F.~Andreev, E.~P.~Bashkin, Zh.~Eksp.~Teor.~Fiz. {\bf 69},
319 (1975) [Sov.~Phys.~JETP {\bf 42}, 164 (1975)].

\bibitem{Volosniev} A.~G.~Volosniev, H.-W.~Hammer, and N.~T.~Zinner, Phys.~Rev.~A {\bf 92}, 023623 (2015).

\bibitem{Werner} F.~ Werner and Y.~Castin, Phys.~Rev.~A {\bf 86}, 053633 (2012).

\bibitem{Kroiss} P.~Kroiss, M.~Boninsegni, and L.~Pollet, Phys.~Rev.~B {\bf 93}, 174520 (2016).

\bibitem{Mora09} C.~Mora and Y.~Castin, Phys.~Rev.~Lett. {\bf 102}, 180404 (2009).
\bibitem{Astrakharchik_etal} G.~E.~Astrakharchik, J.~Boronat, I.~L.~Kurbakov, Yu.~E.~Lozovik, and F.~Mazzanti, Phys.~Rev.~A {\bf 81}, 013612 (2010).

\bibitem{Salasnich} L.~Salasnich, Phys.~Rev.~Lett. {\bf 118}, 130402 (2017).



\end{thebibliography}


\end{document}